\documentclass[12pt]{article}
 
\usepackage{epsfig}
\usepackage{amsmath}
\usepackage{amssymb}
\usepackage{theorem}

\newcommand{\Mol}{{\cal M}} 
\newcommand{\Rea}{{\cal R}} 

\newcommand{\reactsto}{ \to}   

\newcommand{\INR}[2]{(#1 \reactsto #2) \in \Rea}   

\newcommand{\Pow}[1]{{\cal P}(#1)}   
\newcommand{\PowM}[1]{{\cal P}_M(#1)}   

\newcommand{\GO}{G} 

\newcommand{\GSM}{G_{SM}} 

\newcommand{\GCL}{G_{CL}} 
\newcommand{\GMM}{G_{MM}} 
\newcommand{\GSO}{G_{SO}} 

\newcommand{\X}{X} 

\newcommand{\fmol}{\phi}   

\newcommand{\thresh}{\Theta} 

\newcommand{\allCL}{{\cal O}_{CL}}  
\newcommand{\unionCL}{\sqcup_{CL}} 
\newcommand{\intersecCL}{\sqcap_{CL}} 

\newcommand{\allSM}{{\cal O}_{SM}}  
\newcommand{\unionSM}{\sqcup_{SM}} 
\newcommand{\intersecSM}{\sqcap_{SM}} 

\newcommand{\unionSO}{\sqcup_{SO}} 
\newcommand{\intersecSO}{\sqcap_{SO}} 

\newcommand{\allMM}{{\cal O}_{MM}}  
\newcommand{\unionMM}{\sqcup_{MM}} 
\newcommand{\intersecMM}{\sqcap_{MM}} 

\newcommand{\allO} {{\cal O}} 
\newcommand{\unionO}{\sqcup} 
\newcommand{\intersecO}{\sqcap} 

\newcommand{\numberof}[2]{\#(#1 \in #2)} 

\newcommand{\stochmat}{\mathbf{M}}      

\newcommand{\systema}{{\it general reaction system} }
\newcommand{\systemas}{{\it general reaction systems} }
\newcommand{\systemb}{{\it consistent reaction system} }
\newcommand{\systembs}{{\it consistent reaction systems} }
\newcommand{\systemc}{{\it reactive flow system} }
\newcommand{\systemd}{{\it catalytic flow system} }
\newcommand{\systemds}{{\it catalytic flow systems} }

\newcommand{\SystemA}{General Reaction System }
\newcommand{\SystemB}{Consistent Reaction System }
\newcommand{\SystemC}{Reactive Flow System }
\newcommand{\SystemD}{Catalytic Flow System }

\newcommand{\RR}{{\mathbb  R}} 

\theorembodyfont{\rmfamily}
\theoremheaderfont{\itshape}

\newtheorem{defi}{Definition} 
\newtheorem{theorem}{Theorem}
\newtheorem{lemma}{Lemma} 
\newtheorem{example}{Example}

\newcommand{\Def}[2]{\vspace{-.4cm}\begin{defi}  #2 \end{defi}\vspace{-0.2cm}}
\newcommand{\Lem}[1]{\vspace{-.4cm}\begin{lemma}  #1 \end{lemma}\vspace{-0.2cm}}
\newcommand{\Example}[2]{\begin{example}  #2 \end{example}}

\newcommand{\aidx}[1]{{}}
\newcommand{\sidx}[1]{{}}

\newcommand{\df}[1]{{\em #1}}

\newcommand{\beq}{\begin{equation}}
\newcommand{\eeq}{\end{equation}}

\newcommand{\bsy}{\begin{center}\begin{tabular}{|l|l|p{12cm}|} \hline }
\newcommand{\esy}{\hline \end{tabular}\end{center}}

\newcommand{\ben}{\begin{enumerate}}
\newcommand{\een}{\end{enumerate}}

\renewcommand{\vec}[1]{\mathbf{#1}}

\begin{document}

\noindent

{\bf \large Chemical organization theory: towards a theory of constructive dynamical systems}
\vspace{0.3cm} \\
\noindent
{\bf Peter Dittrich, and Pietro Speroni di Fenizio}\footnote{Both authors
contributed equally.}
\vspace{0.3cm} \\
\noindent
Bio Systems Analysis Group, Jena Centre for Bioinformatics \& 
Department of Mathematics and Computer Science, Friedrich-Schiller-University Jena,
D-07737 Jena
\vspace{0.3cm} \\
\today  
\vspace{0.3cm} \\

{
\bf
Complex dynamical networks consisting of many components
that interact and produce each other are difficult 
to understand, especially, when new components may appear. 
In this paper we outline a theory to deal with such systems.
The theory consists of two parts. The first part introduces
the concept of a chemical organization as a closed
and mass-maintaining set of components. This concept allows
to map a complex (reaction) network to the set of 
organizations,  providing a new view on the system's structure.
The second part connects dynamics with the set of
organizations, which allows to map a movement of
the system in state space to a movement in the set of organizations.
}
\vspace{0.5cm}

Our world is changing, qualitatively and quantitatively.
The characteristics of its dynamics can be as simple as in the
case of a friction-less swinging pendulum, or as complex
as the dynamical process that results in the creative apparition of novel
ideas or entities. We might characterize the nature of a  dynamical process
according to its level of novelty production.
For example, the friction-less swinging pendulum implies
a process where the novelty is only quantitative.
Whereas the process of biological evolution is
highly creative and generates qualitative novelties, which then spread in a
quantitative way.

Fontana and Buss \cite{ac:FB94arrival} called processes and systems that
display the production of novelty, constructive (dynamical)
processes and constructive (dynamical) systems, respectively.
Constructive systems can be found on all levels of scientific
abstraction: in nuclear physics, where the collision of
atoms or subatomic particles leads to the creation of 
new particles; in molecular chemistry, where molecules can
react to form new molecules; or in social systems, where 
communication can lead to new communication \cite{soc:Luh84}.
As a result of a combinatorial explosion, it is easy to create
something that is new, e.g., a molecule or a poem that is unique in the
whole known universe.
Nevertheless, it should be noted that novelty is relative. 
Whether something is considered
to be new or not, depends on what is already there.
Thence it follows that whether a system appears as a
constructive dynamical system  is usually context dependent.

Despite the fact that a large amount of interesting
dynamical processes are constructive, classical
systems theory does not, conveniently, take novelty into account.
Classical systems theory assumes a given static set 
of components. For example, a classical systems analysis 
would first identify all components and their relations; then
would identify a state space with fixed dimensionality, e.g.,
the state of our pendulum could be described
by a two-dimensional vector specifying angle and angular velocity.

The lack of a theory for constructive dynamical systems
has been identified and discussed in detail by Fontana and Buss \cite{ac:FB94arrival}
in the context of a theory for biological organization.
As a solution, they suggested the important concept 
of a (biological) organization as
an operationally  closed and dynamically self-maintaining system.
Taking this idea further, we suggested to divide the theoretical
approach in a static and a dynamic analysis~\cite{ped:SD2002}.
Here, we present the results of our study for general reactions systems, 
integrating stoichiometry and the concept of mass-maintenance in the definition of
organizations, which is a prerequisite for a broad applicability.

\paragraph{Reaction Systems.} 
The theory described herein aims at understanding \df{reaction systems}.
A reaction system consists of reactants, which we refer to as \df{molecules}.
An interaction among molecules that lead to the 
appearance or disappearance of molecules is called a \df{reaction}.
In particular, reaction systems are used to model chemical processes,
but in general, they can be applied to virtually any domain where elements
interact to influence the production of other elements, e.g., 
population dynamics, evolution of language, economy, or social dynamics.

We have to distinguish between a reaction system as an abstract description
of all possible molecules (and their reactions), and a reaction vessel,
which contains concrete instances of molecules from the set of all possible molecules.
In general, the description of a reaction system can be
subdivided into three parts: 
(1) the set of all possible molecules $\Mol$, 
(2) the set of all possible reactions among all the possible molecules $\Rea$,
and (3) the dynamics, which describes how the reactions are applied
to a collection of
molecules inside a reaction vessel.

\subsection*{Static Analysis}

\vspace{0.3cm}
\noindent
{\bf Algebraic Chemistry.} In the first part of the paper, we are only concerned with
the static structure of a reaction system, that is, the molecules
and the reactions. And instead of considering a state (e.g., a concentration vector), we
limit ourself to the analysis of the set of molecules present in that state.
We introduce the concept of an algebraic
chemistry, which is a reaction network, including stoichiometric information,
from which we will derive the organizational structure of the system.

\Def{algebraic chemistry}{
Given a set $\Mol$  of elements (called molecules)
and a set of reaction rules given
by the relation $\Rea: \PowM \Mol \times \PowM \Mol$.
We call the pair $\langle \Mol, \Rea \rangle$ an \df{algebraic chemistry}.

}

$\PowM C$ is the set of all multisets with elements from $C$.
A multiset differs from a set in the fact that the same element can appear
more than one time. 
The frequency of occurrence of an element $a$ in a multiset
$A$ is denoted by $\numberof a A$.
For simplicity, we adopt a notion from chemistry to write 
reaction rules.
Instead of writing $(\{ s_1, s_2, \dots, s_n \} ,  
\{ s'_1, s'_2, \dots, s'_{n'} \} ) \in \Rea$
we write:
$  s_1 + s_2 + \dots + s_n \reactsto s'_1 + s'_2 + \dots + s'_{n'}$ .
Given the left hand side molecules $A = \{ s_1, s_2, \dots, s_n \}$ and 
the right hand side molecules 
$B = \{ s'_1, s'_2, \dots, s'_{n'} \}$, we write
$  \INR A B$ instead of $(A, B) \in \Rea$.
$A \reactsto B$ represents a chemical reaction equation
where $A$ is the
multiset of molecules on the left hand side (also called \df{reactants})
and $B$ the multiset of molecules on the right hand side (also called
\df{products}). 

\vspace{0.3cm}
\noindent
{\bf Input and Output.} 
There are many processes that give
rise to an inflow and outflow, such as, incident sunlight, decaying molecules,
or a general dilution flow. 
In this paper we handle input simply by 
adding a
reaction rule $\emptyset \to a$ to the reaction rules $\Rea$
for every molecule $a$ that is in the inflow.
$\emptyset$ denotes the empty set.
Equivalently, for an output molecule  $a$ 
(e.g., a molecule that is decaying) 
we add the rule $a \to \emptyset$ to the reaction rules $\Rea$.

\paragraph{Semi-Organization.} In classical analysis, we study the movement
of the system in state space. Instead, here, we consider the movement from one set
of molecules to another. As in the classical analysis of the dynamic of
the system, where fixed points and attractors are considered more important
than other states, some sets of molecules are more important than others.
In order to find those sets, we introduce some
properties 
that define them, namely: closure, self-maintenance,
semi-organization, mass-maintenance, and finally the organization.

All definitions herein refer to an algebraic chemistry $\langle \Mol,
  \Rea \rangle$. The first requirement, called closure, assures
that a set of molecules contains all molecules that can be
produced by reactions among those molecules.

\Def{closed set}{
  A set $C \subseteq \Mol$ is \df{closed}, 
  if for all reactions $\INR A B$, with 
  $A$ a multiset of elements in $C$  ($A \in \PowM C$),
  $B$ is also a multiset of elements in $C$ ($B \in \PowM C$). 
}

Given a set $S \subseteq \Mol$, we can always generate
its closure $\GCL(S)$ according to the following definition:

\Def{generate closed set}{
Given a set of molecules $S \subseteq \Mol$,
we define $\GCL(S)$ as the smallest closed set
$C$ containing $S$. We say that $S$ \df{generates the closed set} 
$C = \GCL(S)$ and we call $C$ the \df{closure} of $S$. 
}

The closure implies a union and an intersection operator on
the closed sets $\allCL$ of an algebraic chemistry.
Given two closed sets $U$ and $V$,  the
closed set generated by their union ($U \unionCL V$) and intersection
 ($U \intersecCL V$) is defined as:
 $U \unionCL V      \equiv  \GCL(U \cup V)$, and  
 $U \intersecCL V   \equiv  \GCL(U \cap V)$, respectively.
Trivially, closed sets form a lattice
$\langle \allCL, \unionCL, \intersecCL \rangle$,
which is a common algebraic structure
(a poset in which any two elements have a greatest lower bound
and a least upper bound, see supl. mat.).
The property of closure is important, because the 
closed set represents the largest possible set that
can be reached from a given set of molecules.
Furthermore a set that is closed cannot generate 
new molecules and is in that sense more stable than
a set that is not closed.

The next important property, called self-maintenance,
assures that every molecule that is used-up within a set, is
produced within that set.
\Def{self-maintaining set}{
  A set of molecules  $S \subseteq \Mol$
  is called \df{self-maintaining}, if all 
  molecules $s \in S$ that are used-up within $S$ 
  are also produced within that set $S$.
} 
More precisely, we say that a molecule $k \in \Mol$ is \df{produced}
within a set $C \subseteq \Mol$, if there exists a reaction $(A \to B)$, with 
$A \in \PowM C$, 
and $\numberof k A < \numberof k B$.
In the same way, we say that a molecule $k \in C$ is \df{used-up} 
within the set $C$, if, within the set $C$,  
there is a reaction  $(A \to B)$ with 
$A \in \PowM C$, 
and $\numberof k A > \numberof k B$.

Taking closure and self-maintenance together, we arrive at
the important concept of a semi-organization.

\Def{semi-organization}{
A \d{semi-organization} $O \subseteq \Mol$
is a set of molecules that is closed and self-maintaining.
}

\paragraph{Organization.}
In a semi-organization, all molecules that are used-up are produced;
but, this does not guarantee that the total amount of mass can be maintained.
A small, but important example is the reversible reaction in a flow reactor:
$\Mol = \{a, b \}, 
\Rea = \{ a \reactsto b, 
          b \reactsto a , 
          a \reactsto \emptyset,
          b \reactsto \emptyset
\}$, where both molecules also decay. $O = \{ a, b \}$ is a semi-organization,
because the set is closed, $a$ is produced by the reaction $b \reactsto a$,
and $b$ is produced by the reaction $a \reactsto b$. But, obviously, the
system $\{ a, b\}$ is not stable, because both molecules decay and are not sufficiently
reproduced, so that they will finally vanish.
The solution to this problem is to consider the overall ability 
of a set to maintain its total mass. We call such sets mass-maintaining:

\Def{mass-maintaining}{
Given an algebraic chemistry $\langle \Mol, \Rea \rangle$
with $m = |\Mol|$ molecules and $n = |\Rea|$ reactions, and let 
$\stochmat = (m_{i,j})$ be the ($m \times n$) stoichiometric matrix implied
by the reaction rules $\Rea$, where $m_{i,j}$ denotes the number of molecules
of type $i$ produced in reaction $j$. 
  A set of molecules $C \subseteq \Mol$ is called \df{mass-maintaining}, 
  if there exists a flux vector  $\vec v \in \RR^n$ such that the three following
  condition apply: (1) for all  reactions $(A \to B)$ with $A \in \PowM C$ the
  flux $v_{(A \to B)} > 0$; (2) for all reactions  $(A \to B)$ with 
  $A \notin \PowM C$, $v_{(A \to B)}=0$; and 
  (3) for all molecules $i \in C$, 
  $f_i \geq 0$ with $ (f_1, \dots,  f_m) = \stochmat \vec v$.
}
$v_{(A \to B)}$ denotes the element of $\vec v$ describing
the flux of reaction $A \reactsto B$. 
For the example above, the stoichiometric matrix becomes 
$\stochmat = ((-1, 1)$, $(1, -1)$, $(-1, 0)$, $(0, -1))$, and we can see
that there is no positive flux vector $\vec v \in \RR^4$, such that
$\stochmat \vec v \geq \vec 0$. In fact, only the empty 
semi-organization $\{ \}$ is mass-maintaining.
In case $a$ and $b$ would not decay, $\Rea = \{ a \reactsto b, b \reactsto a
\}$, the set $\{a , b\}$ would be (as desired) mass-mainatining, because threre is
a flux vector, e.g., $\vec v = (1.0, 1.0)$, such that $\stochmat \vec v = \vec
0 \geq \vec 0$ with  $\stochmat = ((-1, 1), (1, -1))$.
Now, closure and mass-maintenance together lead to the central definition
of this work:

\Def{organization}{
  A set of molecules $O \subseteq \Mol$ that is closed and
  mass-maintaining is called an \df{organization}.
}

An organization represents an important combination of molecular species,
which are likely to be observed in a reaction vessel on the long run.
A set of molecules that is not closed or not mass-maintaining would not
exists for long, because new molecules can appear or some molecules
would vanish, respectively.
The condition ``mass-maintaining'' is stronger and more difficult to
compute than the condition ``self-maintaining''. In fact, the former implies
the latter, because in a mass-maintaing set, a molecule $i$ that is used-up within
that set must also be produced within that set in order to achieve a 
non-negative production rate $f_i$. Thus we can say: 
\Lem{
Every organization is a semi-organization.
}%

\Example{Four Species}{In order to illustrate the new concepts, we consider
a small example where there are just four molecular species 
$\Mol = \{a, b,c,d \}$, which react according to the following reaction
rules $\Rea = \{ 
a + b \reactsto a + 2b, 
a + d \reactsto a + 2d,
b + c \reactsto 2c,
c \reactsto b,
b + d \reactsto c,
b \reactsto \emptyset, c \reactsto \emptyset, d \reactsto \emptyset 
\}$, where $b,c,d$ decay spontaneously and $a$ is a permanent catalyst.
Although the reaction system is small, its organizational structure
is already difficult to see when looking at the rules or
their graphical representation (Fig.~1).
In Fig.~1, all 16 possible sets of molecules are shown as a lattice.
There are 9 closed sets, 8 self-maintaining sets, 7 mass-maintaining sets, and
7 semi-organizations, 6 of which are organizations. The organizations
are the only combination of molecules that can reside in a
reaction vessel for a long time. 
We can also see immediately that a reactor containing 
$\{b,c,d\}$ must have a transient dynamics ``down''
where molecular species are lost, and that a reactor containing
$\{ a, b, d \}$ will have a transient where a new molecular species
will appear, and so on... 
}

\subsection*{Different reaction systems}

Finding all organizations of a general reaction system
appears to be computationally difficult. 
One approach is to  find the
semi-organizations first, and then check, which of them are also mass-maintaining.
The property of the set of organizations and  semi-organizations 
depends strongly on the type of system studied. 
We will present here four types of systems and analyze their
properties with respect to the newly introduced concepts. 

\paragraph{\SystemD.} 

In a \systemd all molecules are used-up by first-order reactions
of the form $\{ k \} \reactsto \emptyset$ (dilution) and there
is no molecule used-up by any other reaction. 
So, each molecule $k$ decays spontaneously, 
or equivalently, is removed by a dilution flow.
Apart from this, each molecule can appear only as a catalyst (without being used-up).
Examples of \systemd are the replicator equation\cite{ac:SS83}, 
the hypercycle \cite{ac:ES77}, the more general 
{\it catalytic network equation} \cite{ac:SFM93}, or AlChemy \cite{ac:Fon92}. 
Furthermore some
models of genetic regulatory networks %
and social system \cite{ped:DKB2002} are \systemds.
 
\Lem{
In a \systemd, every semin-organization is an organization.
}%

In a \systemd we can easily check, whether a set $O$ is an organization by
just checking whether it is closed and whether each molecule in that
set is produced by that set.
Furthermore,  given a set $A$, we can always generate an organization by
adding all molecules produced by $A$ until $A$ is closed and than
removing molecules that are not produced until $A$ is self-maintaining.
With respect to the intersection and union of (semi-)organizations
the set of all (semi-) organizations of a \systemd forms an 
algebraic lattice (see below), which has already been noted
by Fontana and Buss \cite{ac:FB94arrival}.

\paragraph{\SystemC.} 

In a  \systemc   all molecules are used-up by first-order reactions
of the form $\{ k \} \reactsto \emptyset$ (dilution).
But as opposed to the previous system, we allow arbitrary additional reactions
in $\Rea$. 
This is a typical situation for
chemical flow reactors or bacteria that grow \cite{bio:PK2004}. %
In a \systemc,  semi-organizations are not necessarily organizations, 
as in the reversible reaction example shown before.
Nevertheless, both, the semi-organizations and the organizations form a
lattice  $\langle \allO, \unionO, \intersecO \rangle$,
Moreover, the union ($\unionO$) and intersection  ($\intersecO$)
of any two organizations is an organization
(see below). 

\paragraph{\SystemB.} 
In a \systemb there are two types of molecules:  persistent molecules $P$ and
non-persistent molecules. All non-persistent molecules 
$ k \in \Mol \setminus P$ are 
used-up (as in the two systems before) by first-order reactions
of the form $\{ k \} \reactsto B$ with $k \notin B$; whereas 
a persistent molecule $p \in P$ is not used-up by any reaction at all.
An example of a \systemb is Example~2, where $a$ is a persistent molecule.
The \systemb is the most general of the four systems where
the semi-organizations and organizations always form a lattice,
and where the generate organization operator can  properly be defined (see below). 
As in a \systemc, not all semi-organizations are organizations.

\paragraph{\SystemA.} 

A \systema consists of arbitrary reactions. 
Examples are planetary atmosphere chemistries \cite{ac:YD99}.
In a \systema, neither the set of  organization nor the set of  semi-organizations 
necessarily form a lattice.
Because of the lack of properties there is an actual difficulty in
studying those systems. The analysis gets much easier, if we are able
to transform the system into something else, e.g., a \systemb.
This could be done by introducing a small outflow for each molecule.

\paragraph{Common Properties of \SystemB}

Consistent reaction systems (including \systemc and \systemd) possess
some comfortable properties that allow us to present a series of
useful definitions and lemmas. 
In a \systemb, given a set of molecules $C$, 
we can uniquely {\em generate} a self-maintaining set, a
semi-organization, a mass-maintaining set, and an organization
in a similar way as we have generated a closed set.
And like for closed sets, we can define the union and intersection
on self-maintaining sets, semi-organizations, mass-maintaining sets,
and organizations, respectively.  
Furthermore, each, the self-maintaining sets, semi-organizations, 
mass-maintaining sets,
and organizations form a lattice together 
with their respective union and intersection operators.
This does not generalize to \systema, because
for a \systema we cannot uniquely generate a self-maintaining set, a
semi-organization, a mass-maintaining set, nor an organization
as in the case of a \systemb.

\Def{generate self maintaining set}{
Given a set of molecules $C \subseteq \Mol$,
we define $\GSM(C)$ as the biggest self-maintaining set
$S$ contained in $C$. We say that $C$ 
\df{generates the self-maintaining set} $S = \GSM(C)$. 
}
In order to calculate the self-maintaining set generated by $C$,
we remove those molecules 
that are used-up and not produced within $C$, 
until all molecules used-up are also produced, and thus
reaching a self-maintaining set.
The operator $\GSM$ (generate self-maintaining set) implies
the union $\unionSM$ and intersection $\intersecSM$ on self-maintaining sets:
Given two self maintaining sets $S_1$ and $S_2$,  the
self-maintaining sets generated by their union ($S_1 \unionCL S_2$) and intersection
($S_1 \intersecCL S_2$) are defined as:
 $S_1 \unionSM S_2      \equiv  \GSM(S_1 \cup S_2)$, and  
 $S_1 \intersecSM S_2   \equiv  \GSM(S_1 \cap S_2)$, respectively.
And as already mentioned, in a \systemb, 
$\langle \allSM, \unionSM, \intersecSM \rangle$ forms a lattice,
where $\allSM$ is the set of all self-maintaining sets of an algebraic chemistry.
Finally note that, if $S$ is self-maintaining, its closure
$\GCL(S)$ is self-maintaining, too (in \systembs).

There are many ways in which we can generate a semi-organization from a
set. We will present here the simplest one, which implicitly
assumes that molecules are produced quickly and vanish slowly.
This assumption leads to the largest possible semi-organization
generated by a set:

\Def{generate semi-organization}{
Given a set of molecules $C \subseteq \Mol$,
we define $\GSO(C)$ as $\GSM(\GCL(S))$.
We say that $C$ 
\df{generates the semi-organization} $O = \GSO(C)$. 
}

In the same way as before, the generate semi-organization operator  $\GSO(C)$ implies 
the union  $\unionSO$ and intersection $\intersecSO$ on semi-organizations,
namely 
$O_1 \unionSO O_2      \equiv  \GSO(O_1 \cup O_2)$, and  
$O_1 \intersecSO O_2   \equiv  \GSO(O_1 \cap O_2)$, respectively,
which implies the lattice of semi-organizations.

\Def{generate mass-maintaining set}{
Given a set of molecules $C \subseteq \Mol$,
we define $\GMM(C)$ as the biggest mass-maintaining set
$S$ contained in $C$. We say that $C$ 
\df{generates the mass-maintaining set} $S = \GMM(C)$. 
}
For \systembs,  $\GMM(C)$ is always defined, 
because the union ($\cup$)
of two mass-maintaining sets is mass-maintaining; and further,
every set is either mass-maintaining, or it contains a unique biggest
mass-maintaining set. Thus from every set we can generate a mass-maintaining
set.
Note that mass-maintaining sets are also self-maintaining,
$\GMM(\GSM(S)) \equiv \GMM(S)$, which is a useful property, 
because  $\GSM(S)$  is easier to compute.
As usual, the union $\unionMM$ and intersection $\intersecMM$ of mass-maintaining sets
$S_1, S_2$ are defined as
$S_1 \unionMM S_2      \equiv  \GMM(S_1 \cup S_2)$,
$S_1 \intersecMM S_2  \equiv  \GMM(S_1 \cap S_2)$, respectively.
Thus also the set of all mass-maintaining sets  $\allMM$ forms a lattice
$\langle \allMM, \unionMM, \intersecMM \rangle$.
If $S$ is mass-maintaining, its closure $\GCL(S)$ is mass-maintaining, too
(again, not valid for \systemas).

Finally, in \systembs, we can also generate uniquely an organization,
(here, again, the largest organization that can be generated from a set) 
according to the following definition:

\Def{generate  organization}{
Given a set of molecules $C \subseteq \Mol$,
we define $\GO(C)$ as $\GMM(\GCL(C))$.
We say that $C$ 
\df{generates the organization} $O = \GO(C)$. 
}

Equivalently $\GO(C) = \GMM(\GSM(\GCL(C)))$, which allows to
compute the organization generated by a set more easily in three steps.
Following the same scheme as before,
the union $\unionO$ and intersection $\intersecO$ of two organizations
$U$ and $V$ is defined as the
organization generated by their set-union  and set-intersection:
$U \unionO V      \equiv  \GO(U \cup V)$, 
$U \intersecO V   \equiv  \GO(U \cap V)$,
respectively. 
Thus, for \systemb, also the set of all organizations  $\allO$ forms a lattice
$\langle \allO, \unionO, \intersecO \rangle$. This important fact should be
emphasized by the following lemma:

\Lem{
  Given an algebraic chemistry $\langle \Mol, \Rea
  \rangle$ of a \systemb  and all its organizations $\allO$,
  then  $\langle \allO, \unionO, \intersecO \rangle$ is a
  lattice.
}%

Knowing that the semi-organizations and organizations form a lattice, and that we
can uniquely generate an organization for every set, 
is a useful information. In order to find the
whole set of organizations, it is impractical just to check all the possible sets
of molecules.
Instead,  we can start by computing the lattice of semi-organizations,
and then test only those sets for mass-maintenance. 
Furthermore, if the semi-organizations form a lattice, we can start with
small sets of molecules and generate their semi-organizations,
while the $\unionSO$ operator can lead us to the more
complex semi-organizations.

As a summary, from a practical point of view, we 
calculate first the set of semi-organizations.
If our system is a \systemd, we automatically obtain the lattice
of organizations. Otherwise we have to check for each semi-organization whether
it is mass-maintaining or not. If we have a \systemb, then
we are assured to obtain a lattice, where there is a unique smallest and 
largest organization, and where we can easily obtain the intersection
and union of two organizations from the graphical representation of
the lattice (see examples).
For a \systema, the set of organizations does not 
necessarily form a lattice. 
Nevertheless, this set of organizations represent
the \df{organizational structure} of the reaction network, which can
be visualized and which can provide a new view
on the dynamics of the system by mapping  the
movement of the system in state space to 
a movement in the set of organizations, as will be
shown in the following section.

%
%

%


\subsection*{Dynamic Analysis}

The static theory  
deals with  molecules $\Mol$ and their reaction rules $\Rea$, but not
with the evolution of the system in time.
To add dynamics to the theory, we have
to formalize the dynamics of a system.
In a very general approach, 
the \df{dynamics} is given  by a \df{state space} $X$ and 
a formal definition (mathematical or algorithmical) that
describes all possible  movements in $X$.
Given an initial state $\vec x_0 \in X$,   
the formal definition describes how the state changes over time.
For simplicity, we assume a deterministic dynamical process, which
can be formalized  by a phase flow 
$(X, (T_t)_{t \in \RR})$ where 
   $(T_t)_{t \in \RR}$ is a one-parametric group
   of transformations from $X$. $T_t(\vec x_0)$ denotes the state at time $t$ of
   a system that has been in state $\vec x_0$ at $t=0$.
\paragraph{Connecting to the Static Theory}
\label{sec:conn-stat-theory}

A state $\vec x \in \X$ represents the state of a reaction vessel 
that contains molecules from $\Mol$.
In the static part of the theory we consider just the set
of molecular species present in the reaction vessel, but not
their concentrations, spatial distributions, velocities, and so on.

Now, given the state $\vec x$ of the reaction vessel, we need a function
that maps uniquely this state to the set of molecules present.
Vice versa, given a set of molecules $A \subseteq \Mol$,
we need to know, which states from $\X$ correspond to this set of molecules.
For this reason we introduce a mapping $\fmol$ called \df{abstraction},
from $\X$ to $\Mol$, which maps a state of the system to the set of molecules
that are present in the system being in that state.
We require this mapping to be monotonically increasing on the number of molecules. 
In other words, if we have two states
$\vec x_1$ and $\vec x_2$, and the concentration of the molecule $m_1$ is higher in
$\vec x_2$ than in $\vec x_1$; and if $m_1 \in \fmol(\vec x_1)$ then 
$m_1 \in \fmol(\vec x_2)$. The exact mapping can %
be defined precisely later, depending on the state space, on the dynamics,
and on the actual application.

The concept of \df{instance} is the opposite of the concept of abstraction.
While $\fmol(\vec x)$
denotes the molecules represented by the state $\vec x$, an
instance $\vec x$ of a set $A$ is a state where exactly the molecules
from $A$ are present according to the function $\fmol$.

\Def{instance of $A$}{
We say that a state $\vec x \in X$ is an instance of $A \subseteq \Mol$,
iff $\fmol(\vec x) = A$
}

In particular, we can define an instance of an organization $O$ 
(if $\fmol(x) = O$) and an instance of a generator of $O$ 
(if $\GO(\fmol(x)) = O$).
Loosely speaking we can
say that $\vec x$ \df{generates organization} $O$.
Note that a state $\vec x$ of a  \systemb, \systemc, and \systemd 
is {\em always} an instance of a generator 
of one and only one organization $O$. 
This leads to the important observation that
a lattice of organizations partitions the state space $\X$,
where a partition $\X_O$ implied by organization $O$ is
defined as the set of all instance of all generators of $O$:
$\X_O = \{\vec x \in \X |  \GO(\fmol(\vec x)) = O \}$.
Note that as the system state evolves over time,
the organization $G(\fmol(\vec x(t)))$ 
generated by $\vec x(t)$ might change (see below, Fig.~2~and~3).

%

\paragraph{Fixed Points are Instances of Organizations.}

Now we describe a theorem that relates
fixed points to organizations, and by doing so, underlining
the relevancy of organizations.
We will show that, 
given an ODE of a form that is commonly used to describe 
the dynamics of reaction systems, 
every fixed point of this ODE is an instance of an organization.
We therefore assume in this section
that $\vec x$ is a concentration vector
$\vec x = (x_1, x_2, \dots, x_{|\Mol|})$, $\X = \RR^{|\Mol|}, x_i \geq 0$ 
where $x_i$ denotes 
the concentration of molecular species $i$ in the reaction vessel, 
and $\Mol$ is finite.
The dynamics is given by an ODE of the form 
$\dot {\vec x} = \stochmat \vec v(\vec x)$ where 
$\stochmat$ is the stoichiometric matrix implied by the 
algebraic chemistry $\langle \Mol, \Rea \rangle$ (reaction rules).
$\vec v(\vec x) = (v_1(\vec x), \dots, v_n(\vec x)) \in \RR^{|\Rea|}$ is a flux vector
depending on the current concentration $\vec x$, where $|\Rea|$ denotes
the number of reaction rules.
A flux $v_j(\vec x)$ describes the rate of a particular reaction $j$.
For the function $v_j$ we require only that $v_j(\vec x)$ is positive,
if and only if the molecules on the left hand side of the reaction $j$ are
present in the state $\vec x$, and otherwise it must be zero. 
Often it is also assumed that  $v_j(\vec x)$ increases monotonously,
but this is not required here.  %
Given the dynamical system as $\dot {\vec x} = \stochmat \vec v(\vec x)$, 
we can define the abstraction of a state
$\vec x$ formally  by using
a (small) threshold $\thresh \geq 0$ 
such that all fixed points have
positive coordinates greater than $\thresh$.

\Def{abstraction}{
  \label{def:abstraction}
  Given a dynamical system $\vec {\dot x} = f(\vec x)$ and
  let $\vec x$ be a state in $\X$, then the abstraction $\fmol(\vec x)$ is
  defined by
  \begin{equation}
    \label{eq:fmol}
    \fmol ( \vec x ) = \{ i | x_i > \thresh, i \in \Mol \}, \quad 
    \fmol: \X \to \Pow{\Mol}, \quad \thresh \geq 0     
  \end{equation}
  where $x_i$ is the concentration of molecular species $i$ in state $\vec x$,
  and $\thresh$ is a threshold chosen such that it is smaller than any positive
  coordinate of any fixed point of  $\vec {\dot x} = f(\vec x), x_i \geq 0$.
}

Setting $\thresh=0$ is a safe choice, because in this case $\fmol$
 always meets the definition above. But for practical reasons,
it makes often sense to apply a positive threshold greater zero,
e.g., when we take into consideration that the number of molecules
in a reaction vessel is finite. 

\begin{theorem}
{\bf Hypothesis:} Let us consider a \systema whose
reaction network is given by the algebraic chemistry  $\langle \Mol, \Rea
\rangle$ and whose dynamics is 
given by $\dot {\vec x} = \stochmat \vec v(\vec x) = f(\vec x)$ as defined before.
Let $\vec x' \in \X$ be a fixed point, that is, $f(\vec x')= \vec 0$, 
and let us consider a mapping $\fmol$ as given by Def.~\ref{def:abstraction},
which assigns a set of molecules to each state $\vec x$.
{\bf Thesis:}
$\fmol(\vec x')$ is an organization. 
\end{theorem}

  {\bf Proof:}
   We need to prove that $\fmol(\vec x')$ is closed and
  mass-maintaining: 
  (a) Closure: Let us assume that $\fmol(\vec x')$ is not closed, then
  there exist a molecule $k$ such that $k \notin \fmol(\vec x')$ and
  $k$ is generated by molecules contained in $\fmol(\vec x')$.
  Since, $\vec x'$ is a fixed point, 
    $f(\vec x')= \stochmat \vec v(\vec x') = 0$.
  Now we decompose the stoichiometric matrix $\stochmat$ into
  two matrices $\stochmat^+$ and $\stochmat^-$ separating  all
  positive from the negative coefficients, respectively, such that
  $\stochmat = \stochmat^+ + \stochmat^-$ and  
  $\stochmat^+ \vec v(\vec x') \geq \vec 0$, and
  $\stochmat^- \vec v(\vec x') \leq \vec 0$ (note that 
  $\vec v(\vec x')$ is always non-negative by definition).
  Let $\dot x^+_k$ and  $\dot x^-_k$ be the $k$-th row of 
  $\stochmat^+ \vec v'(\vec x)$ and
  $\stochmat^- \vec v'(\vec x)$, respectively, which represent
  the inflow (production) and outflow (destruction) of molecules
  of type $k$. Note that  
  $\dot x^+_k + \dot x^-_k = \dot x'_k = 0$ (fixed point condition).
  Since we assumed that $k$ is produced by molecules from $\fmol(\vec x')$,
  $\dot x^+_k$ must be positive, $\dot x^+_k > 0$ and thus
  $\dot x^-_k < 0$.
  But this leads to a contradiction. In order to show this
  we have to differentiate the following two cases:
  (i) Assume that $x'_k=0$, then $\dot x^-_k$ must be zero, 
  too (by definition; and intuitively, because a molecule not present cannot
   vanish).
  (ii) Assume that $0 < x'_k \leq \thresh$ ($x'_k> \thresh$ needs not to 
  be considered, because in that case 
  $k$ would be contained in $\fmol(\vec x)$.).
  Because we assumed that $x'_k$ is a coordinate of a fixed point, 
  $\thresh$ must be smaller than $x'_k$, which is again a contradiction.

  (b) Mass-maintaining:  We have to show that $\fmol(\vec x')$ 
  is mass-maintaining.
  Since $\vec x'$ is a fixed point $\stochmat \vec v(\vec x') = \vec 0$, 
  which fulfills condition (3) of the definition of mass-maintaining (Def.~6).
  From the requirements for the flux vector $\vec v$,
  it follows directly that $v_{(A \reactsto B)}(\vec x')>0$ for all 
  $A \in \PowM{\fmol(x')}$, which fulfills condition (1) of Def.~6. 
  Following the same contradictory argument as before in (ii), 
  $x'_k$ must be zero for $k \notin  \fmol(\vec x)$, and therefore
  $v_{(A \reactsto B)}(\vec x')=0$ for $A \notin \PowM{\fmol(\vec x)}$,
  which fulfills the remaining condition (2) of mass-maintaining.
  q.e.d.

From this theorem it follows immediately that a fixed point is an
instance of a closed set, a self-maintaining set, and of a semi-organization.
Let us finally mention that
even if each fixed point is an instance of an organization,
an organization does not necessarily possess a fixed point.
Further note that given an attractor $A \subseteq \X$, 
there exists an organization $O$
such that all points of $A$  are instances of
a generator of $O$.
In fact, it might be natural to suppose 
that all points of an attractor are actually instances of $O$,
 yet it is not clear if this is true for all systems or just for some.

%

\subsubsection*{Movement from Organization to Organization}

%
\paragraph{ODEs and Movement in the Set of Organizations}

 Not all system can be studied using ODEs. In particular a discrete system 
is usually  only {\it approximated} by an ODE.
In a discrete dynamical system, the molecular species that are present in the
reaction vessel can change in time, e.g,  
as the last molecule of a certain type
vanishes. In an ODE instead, this does
not generally happen, where molecules can {\it tend to} zero as
  time {\it tends to} infinity. 
So, even if in reality a molecule disappears, 
in an ODE model it might still be present in a tiny quantity. 
The fact that every molecule ends up being present in (at least) a
  tiny quantity, generally precludes us to notice that the system is actually
  moving from a state where some molecules are present, 
to another state where a
  different set of molecules is present. Yet this is what happens in reality,
  and in this respect, an ODE is a poor approximation of reality. 
A common approach to overcome this problem is to introduce a concentration 
threshold $\thresh$, below which a molecular species is considered not to
be present. We use this threshold in order to define the abstraction 
$\fmol$, which just returns the set of molecules present in a certain state.
Additionally, we might use the threshold to manipulate the numerical
integration of an ODE by setting a concentration to zero, when it falls
below the threshold. In this case, a constructive perturbation has 
to be greater than this threshold.

%

\paragraph{Downward Movement}
\label{sec:movement-down-down}

Not all organizations are stable. %
The fact that there exits a flux vector, such
that no molecule of that organization vanishes,
does not imply that this flux vector can be realized 
when taking dynamics into account.
As a result a molecular species can disappear.
Each molecular species that disappears simplifies the
system. Some molecules can be generated back. But eventually the system can
move from a state that generates organization $O_1$ into a  state that generates
organization $O_2$, with $O_2$ always below $O_1$ ($O_2 \subset O_1$). 
We call this spontaneous movement a \df{downward movement}.

\paragraph{Upward Movement}
\label{sec:movement-up-up}

Moving up to an organization above requires that
a new molecular species appears in the system. 
This new molecular species cannot be produced by a reaction among
present molecules (condition of closure).  
Thus moving to an organization above is more complicated then the
movement down and requires a couple of specifications
that describe how new molecular species enter the system.
Here we assume that new molecular species appear by some sort
of random perturbations or purposeful interference. 
We assume that a small quantity of molecules of that 
new molecular species (or a set of molecular species) suddenly appears.
Often, in practice, the perturbation (appearance of new molecular species) has a
much  slower time scale than the internal dynamics (e.g., chemical reaction kinetics)
of the system. 

\Example{Upward and downward Movement}{
\label{exa:movement-up}
Assume a system with two molecular species $\Mol = \{ a, b \}$ and
the reactions $\Rea = \{ a  \reactsto  2a, b \reactsto 2b, a + b \reactsto a, 
a \reactsto, b \reactsto \}$. All combinations of molecules are organizations,
thus there are four organizations (Fig.~2).
Assume further that the dynamics is governed by
the ODE $\dot x_1 =  x_1 - x_1^2, \dot x_2 = x_2 - x_2^2 - x_1 x_2$ where
$x_1$ and $x_2$ denote the concentration of species $a$ and $b$, respectively.
We map a state to a set of molecules by
using a small, positive threshold $\thresh = 0.1 > 0$ (Def.~\ref{def:abstraction}).
Now, assume that the system is in state $\vec x_0 = (0,1)$ thus in
organization $\{ b \}$.
If a small quantity of $a$ appears (constructive perturbation), 
the amount of $a$ will grow and $b$ will tend to zero.
The system will move upward to organization $\{ a, b \}$ in a transient
phase, while finally moving down and converging to a fixed point in
organization $\{ a \}$. This movement can now be visualized in the lattice
of organizations as shown in Fig.~2.
}

%

\paragraph{Visualizing Possible Movements in the Set of Organizations}

%

In order to display potential movements in the lattice or set of
organizations, we can draw links between
organizations. As exemplified in Fig.~2 and Fig.~3, these links can
indicate possible downward movements (down-link, blue) or upward movements (up-link,
red). A neutral link (black line) denotes that neither the system can 
move spontaneously down, nor can a constructive perturbation move the system
up. Whether the latter is true depends on the definition of ``constructive
perturbation'' applied. For the example of Fig.~3 we defined a constructive
perturbation as inserting a small quantity of {\it one} new molecular species.

The dynamics in between organizations is more complex than this intuitive
presentation might suggest, for example in some cases it is possible to 
move from one organization $O_1$ to an organization $O_2$, with $O_2$ above
(or below) $O_1$
without passing through the organizations in between $O_1$ and $O_2$.
In Fig.~3 this is the case for an upward movement from organization
$\{ a \}$ to organization $\{a,b,c\}$ caused by a constructive perturbation
where a small quantity of $c$ has been inserted. 

%

\paragraph{Organizations in Real Systems}

In preliminary studies \cite{ped:SDZB2000} we have shown
that artificial chemical reaction networks that are based on a 
structure-to-function mapping (e.g., ref.~\cite{ac:FB94arrival,ac:Ban93})
possess a more complex lattice
of organization than networks created randomly. 
From this observation we can already expect that natural
networks possess non-trivial organization structures.
Our investigation of planetary photo-chemistries \cite{ac:YD99} and
bacterial metabolism (which will be published elsewhere), 
revealed lattices of organizations that vanish when the networks are
randomized, indicating a non-trivial structure.
Here, in order to give an impression, we show a lattice
of organizations obtained from a model of the central sugar
metabolism of {\it E.coli} by Puchalka and Kierzek \cite{bio:PK2004}.
The model consists of 92 species and 197 reactions, including gene expression,
signal transduction, transport, and enzymatic activities.
Figure~4 shows the lattice of organizations resulting from the 
original model, ignoring inhibiting interactions.
The smallest organization, $O_1$, contains 76 molecules including
the glucose metabolism and all input molecules.
The input molecules, chosen according to ref.~\cite{bio:PK2004},
include the external food set (Glcex, Glyex, Lacex) and
all promoters (see suppl. material).
Two other organizations, $O_3$ and $O_4$, contain the Lactose and Glycerol
metabolism, respectively. Their union results in the largest
organization $O_5$ that contains all molecules.
In summary, we can conclude, that the organizations found are
biological meaningful, indicating a promising potential for
future applications of this theory.

%

\paragraph{Conclusion}
We presented a general way to define organizations in reaction systems, and
proved a theorem that showed that fixed points are instances of those
organizations. We also investigated the relative structure of the
organizations, opening up the door to visualize the dynamical movement of the
system through organizations. Much work still needs to be done, and we believe
that this line of research is just in its infancy. Reaction systems
with inhibiting interactions need to be investigated, as well as the structure
inside organizations (i.e. the relation between attractive states and
organizations).  In fact the whole issue of movement among sets of molecules
could only be introduced here. The concepts of attractive and stable
organizations, which play
a fundamental role in dynamical systems, has still to be formally defined.
Yet we believe that our work represents a
step forward towards a formal study of constructive dynamical system.
\vspace{0.2cm}

\noindent
{\footnotesize
We thank Florian Centler for providing the {\it E. Coli.} analysis. 
This work was supported by Federal Ministry of Education and Research (BMBF)
Grant~0312704A to Friedrich-Schiller-University Jena.
}

%

%

\newpage
{\bf FIGURE CAPTION}

\vspace{2cm}
Figure~1: Example with four species. Reaction network (a ``2'' means
that two molecules are produced) and graphical
representation of the lattice of all
sets. The vertical position is determined by the set size.
The solid lines depict the lattice of organizations.

\vspace{2cm}

Figure~2: Example of an up-movement caused by a constructive perturbation,
followed by a down-movement. (a) reaction network, 
(b) concentration vs. time plot of a trajectory,
(c) lattice of organizations including trajectory.  

\vspace{2cm}

Figure~3: Lattice of organizations of Example 1, including up-links and down-links.
Furthermore a trajectory is shown starting in organization $\{a,b,c,d\}$
moving down to organization $\{a , b\}$.

\vspace{2cm}

Figure~4: Lattice of organizations of a model of the central sugar
metabolism of {\it E.coli.} \cite{bio:PK2004}. 
In an organization, only names of new molecular species are printed that
are not present in an organization below.

\newpage
 \vspace{2cm}
  \noindent
  \begin{tabular}{p{\linewidth}}
  \centerline{\epsfig{figure=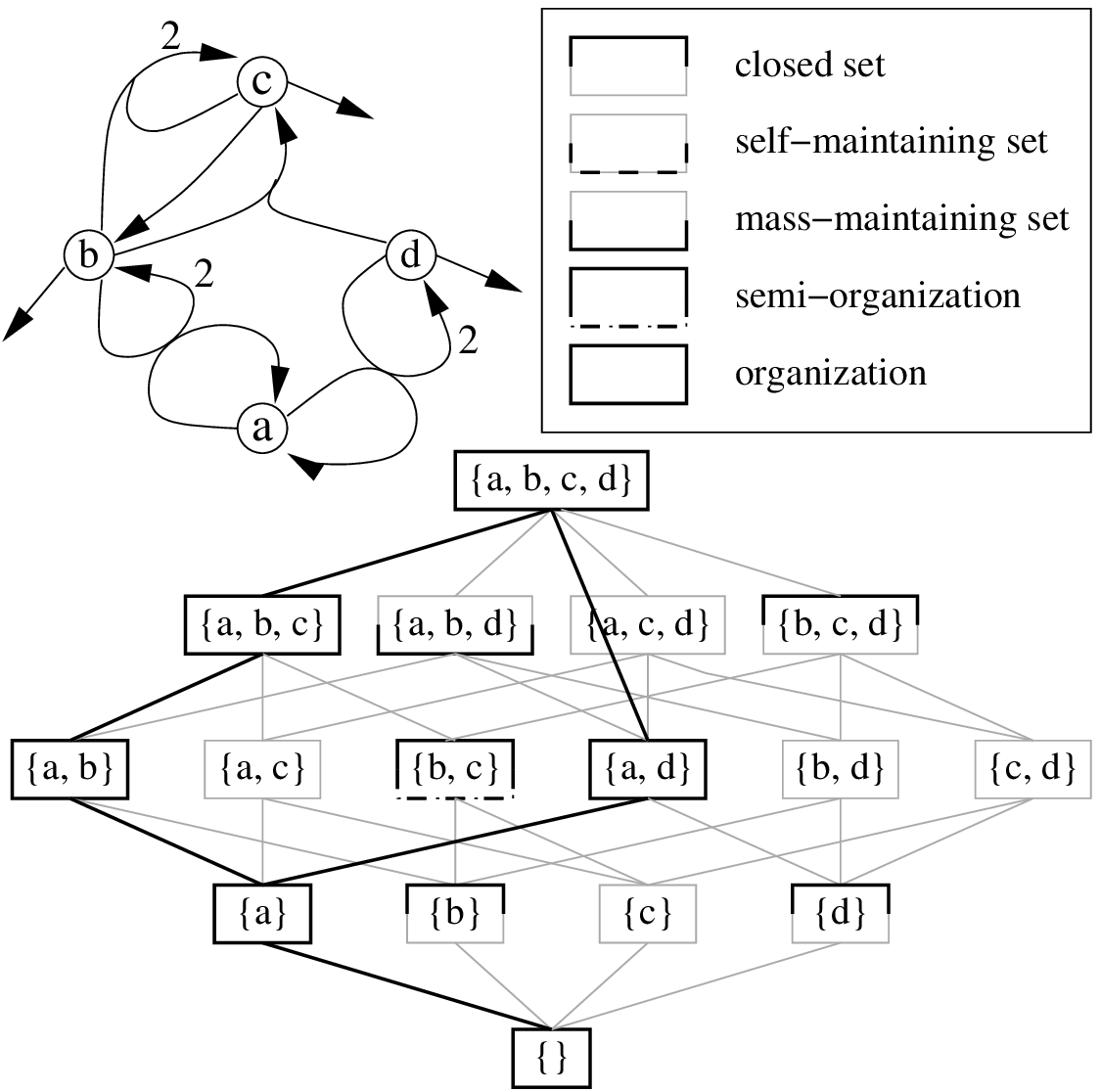,width=10cm}} \\
  \vspace{3cm}

 {Figure~1: Example with four species. Reaction network (a ``2'' means
    that two molecules are produced) and graphical
    representation of the lattice of all
    sets. The vertical position is determined by the set size.
     The solid lines depict the lattice of organizations.
  }
  \end{tabular}
\newpage

\vspace{2cm}
 \noindent
 \centerline{\epsfig{figure=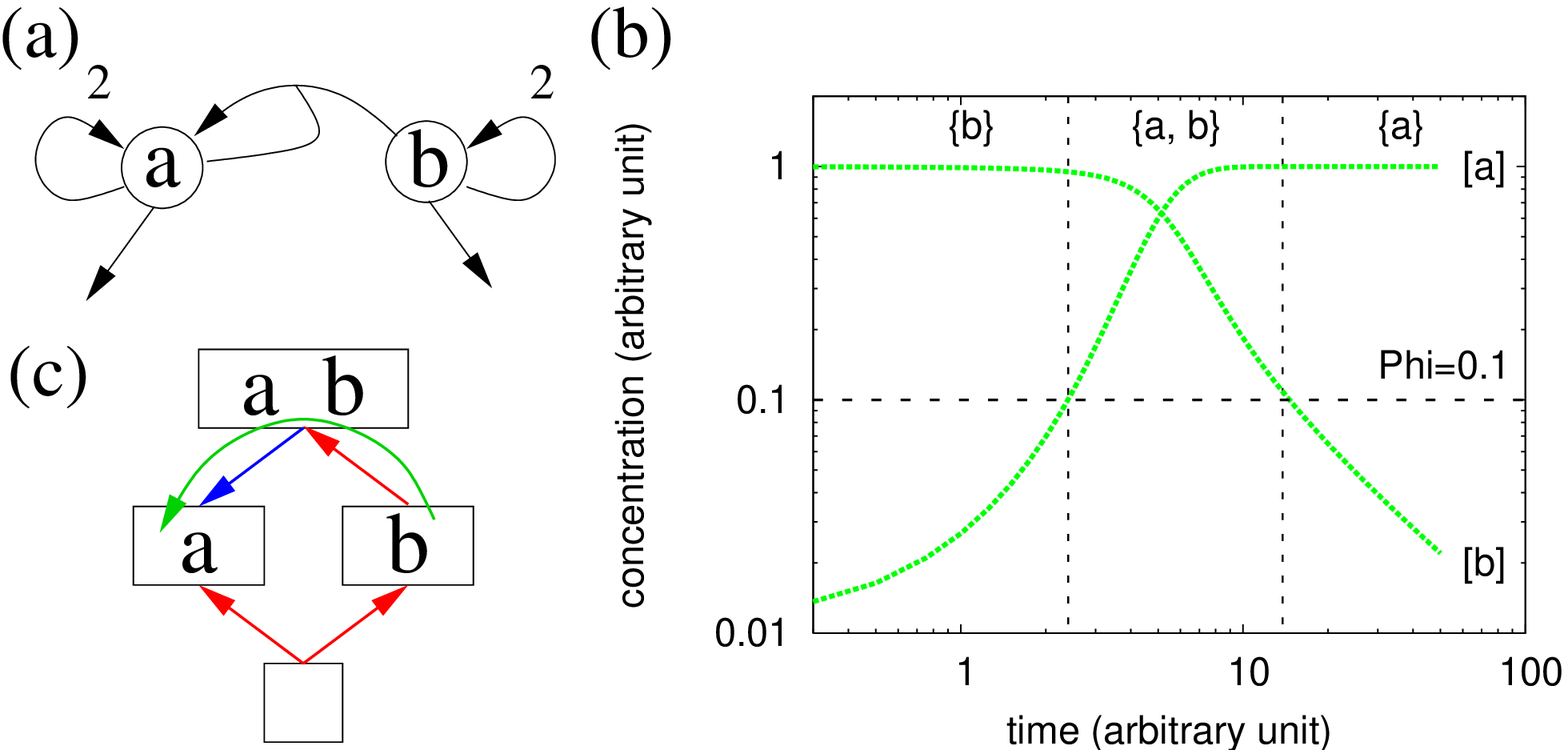,width=10cm}} \\
 \vspace{3cm}

  \noindent
{Figure~2: Example of an up-movement caused by a constructive perturbation,
 followed by a down-movement. (a) reaction network, 
   (b) concentration vs. time plot of a trajectory,
 (c) lattice of organizations including trajectory.  
 }

\newpage
\vspace{2cm}
 \hspace{0.3cm}
 \noindent
 \centerline{\epsfig{figure=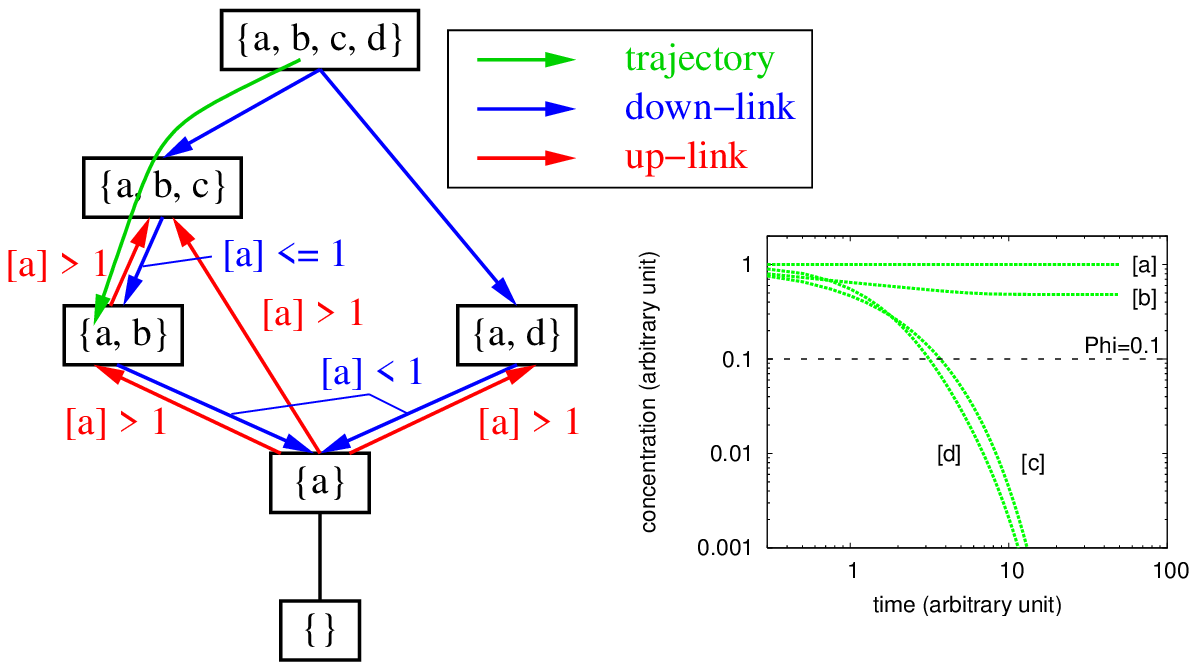,width=10cm}}\\
\vspace{3cm}

  \noindent
 {Figure 3: Lattice of organizations of Example 1, including up-links and down-links.
 Furthermore a trajectory is shown starting in organization $\{a,b,c,d\}$
 moving down to organization $\{a , b\}$.}

\newpage
 \vspace{2cm}
 \hspace{0.3cm}
 \noindent
 \centerline{\epsfig{figure=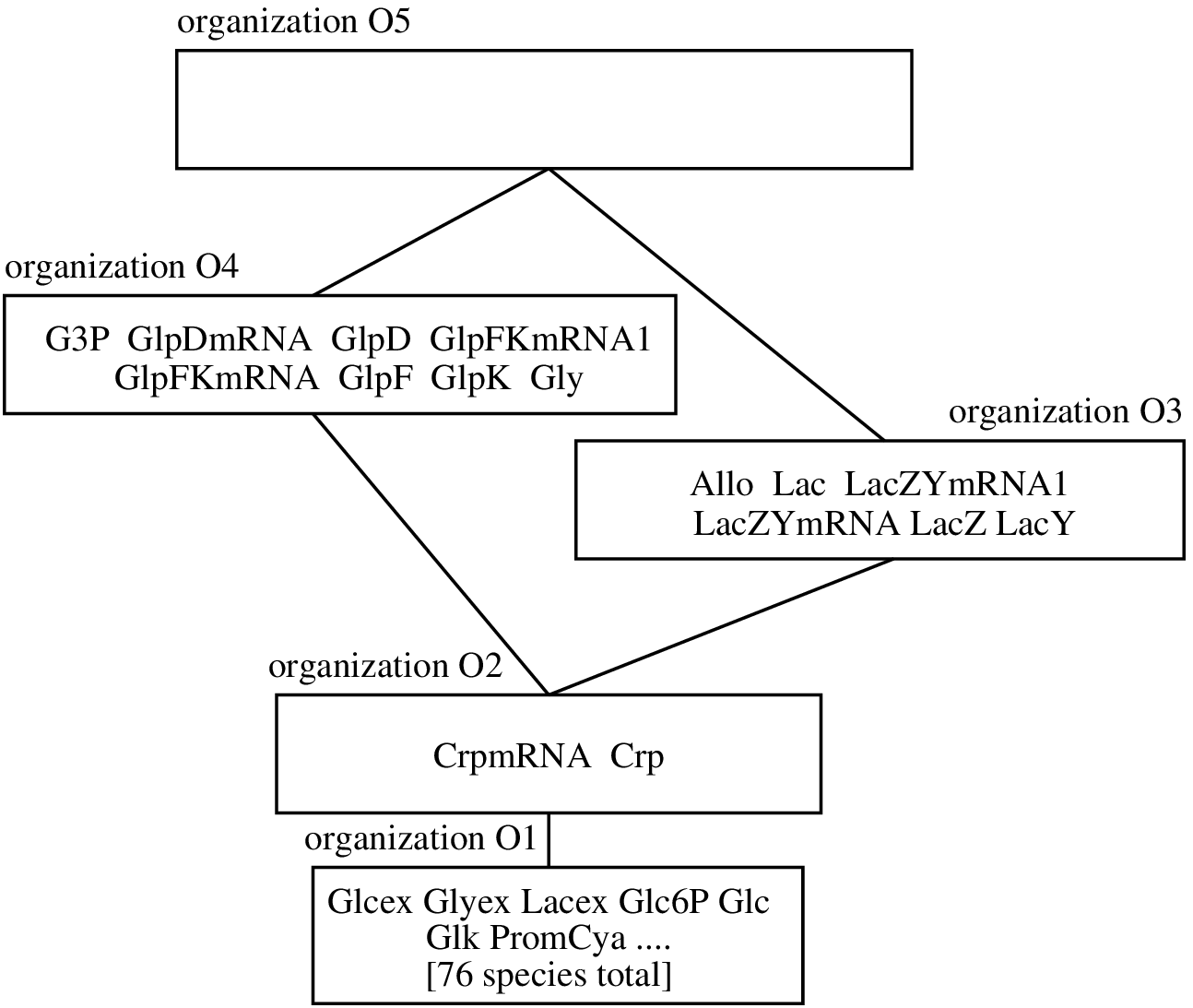,width=10cm}}\\
 \vspace{3cm}

  \noindent
 {Figure 4: Lattice of organizations of a model of the central sugar
 metabolism of {\it E.coli.} \cite{bio:PK2004}. 
 In an organization, only names of new molecular species are printed that
 are not present in an organization below.}
 

\begin{thebibliography}{10}

\bibitem{ac:FB94arrival}
Fontana\aidx{Fontana, Walter}, W \& Buss\aidx{Buss, Leo W.}, L.~W.
\newblock (1994) {\em Bull. Math. Biol.} {\bf 56}, 1--64.

\bibitem{soc:Luh84}
Luhmann, N.
\newblock (1984) {\em Soziale Systeme}.
\newblock (Suhrkamp, Frankfurt a.M.).

\bibitem{ped:SD2002}
Speroni~di Fenizio, P \& Dittrich, P.
\newblock (2002) {\em J. of Three Dimensional Images} {\bf 16},
  160--163.

\bibitem{ac:SS83}
Schuster\aidx{Schuster, Peter}, P \& Sigmund\aidx{Sigmund, Karl}, K.
\newblock (1983) {\em J. Theor. Biol.} {\bf 100}, 533--8.

\bibitem{ac:ES77}
Eigen\aidx{Eigen, Manfred}, M \& Schuster\aidx{Schuster, Peter}, P.
\newblock (1977) {\em Naturwissenschaften} {\bf 64}, 541--565.

\bibitem{ac:SFM93}
Stadler\aidx{Stadler, Peter F.}, P.~F, Fontana\aidx{Fontana, Walter}, W,  \&
  Miller, J.~H.
\newblock (1993) {\em Physica D} {\bf 63}, 378--392.

\bibitem{ac:Fon92}
Fontana\aidx{Fontana, Walter}, W.
\newblock (1992) in {\em Artificial Life II} eds.{} Langton, C.~G, Taylor, C,
  Farmer, J.~D,  \& Rasmussen, S.
\newblock (Addison-Wesley, Redwood City, CA), pp. 159--210.

\bibitem{ped:DKB2002}
Dittrich, P, Kron, T,  \& Banzhaf, W.
\newblock (2003) {\em JASSS}
  {\bf 6}.

\bibitem{bio:PK2004}
Puchalka, J \& Kierzek, A.
\newblock (2004) {\em Biophys. J.} {\bf 86}, 1357--1372.

\bibitem{ac:YD99}
Yung, Y.~L \& DeMore, W.~B.
\newblock (1999) {\em Photochemistry of Planetary Athmospheres}.
\newblock (Oxford University Press, New York).

\bibitem{ped:SDZB2000}
Speroni~di Fenizio, P, Dittrich\aidx{Dittrich, Peter}, P, Ziegler, J,  \&
  Banzhaf\aidx{Banzhaf, Wolfgang}, W.
\newblock (2000) {\em Towards a Theory of Organizations}.
\newblock (GWAL 4, Bayreuth, 5.-7. April, 2000).

\bibitem{ac:Ban93}
Banzhaf\aidx{Banzhaf, Wolfgang}, W.
\newblock (1993) {\em Biol. Cybern.} {\bf 69}, 269--281.

%

\end{thebibliography}
\end{document}